# Highly flexible protein-peptide docking using CABS-dock

Authors: Maciej Paweł Ciemny†, Mateusz Kurcinski†, Konrad Jakub Kozak, Andrzej Kolinski, Sebastian Kmiecik*

Faculty of Chemistry, University of Warsaw, Pasteura 1, 02-093 Warsaw, Poland

† - these authors contributed equally to this paper
* - corresponding author, email: sekmi@chem.uw.edu.pl

**ABSTRACT**

Protein-peptide molecular docking is a difficult modeling problem. It is even more challenging when significant conformational changes that may occur during the binding process need to be predicted. In this chapter, we demonstrate the capabilities and features of the CABS-dock server for flexible protein-peptide docking. CABS-dock allows highly efficient modeling of full peptide flexibility and significant flexibility of a protein receptor. During CABS-dock docking, the peptide folding and binding process is explicitly simulated and no information about the peptide binding site or its structure, is used. This chapter presents a successful CABS-dock use for docking a potentially therapeutic peptide to a protein target. Moreover, simulation contact maps, a new CABS-dock feature, are described and applied to the docking test case. Finally, a tutorial for running CABS-dock from the command line or command line scripts is provided. The CABS-dock web server is available from http://biocomp.chem.uw.edu.pl/CABSdock/

1. INTRODUCTION

Protein-peptide interactions play a predominant role in cell function and they can be found in a variety of signaling pathways involved in cellular localization, immune response or protein expression and degradation. Because of their association with cellular regulatory mechanisms, erroneous protein-peptide interactions are speculated to be pathogenic in a number of diseases (e.g., cancer, autoimmune diseases). The possible applications in biomedical research (targeted drug design) make the understanding of protein-peptide interactions a critical issue for further advances in the field [1, 2]. Characterization of protein-peptide interactions is difficult due to their large complexity and transient and dynamic nature. Despite extensive computational and experimental studies in this area, peptide-mediated cellular regulation mechanisms have not been fully described or understood.

Among computational approaches, molecular docking is commonly used to predict the structure of protein-peptide complexes. Handling large conformational changes during docking is one of the most challenging and important issues in the field [3, 4]. Modeling of protein-peptide interactions usually follows two steps realized by separate protocols: (1) prediction of binding site location on the protein surface [5-8], and (2) local protein-peptide docking (i.e. modeling of the peptide backbone in the binding site) [9-14]. The CABS-dock method [15, 16] unifies these two steps into one efficient docking simulation. In the CABS-dock single simulation run, a fully flexible peptide explores the entire surface of a flexible protein receptor in search for a binding site (no information about the binding site is used). Such high modeling efficiency is achieved thanks to the simulation engine based on the CABS prediction platform [17-19]. Alongside with the Rosetta platform [20], CABS currently offers perhaps the most efficient means for modeling significant conformational changes, successfully tested in protein-peptide on-the-fly docking [21].

This chapter provides a tutorial for the CABS-dock server and for its possible applications. The **Materials** section gives a short description of the CABS-dock methodology, together with the information and instructions required to successfully perform a CABS-dock run. It is followed by the **Methods** section which serves as a step-by-step guide with example docking results and analysis. A description of simulation contact maps, the new CABS-dock feature, is also provided along with the examples of use. Subsequently, possible schemes for incorporating CABS-dock in the multi-stage modeling of protein-peptide interaction are given. Finally, a tutorial how to use the CABS-dock server from the command line or command line scripts is provided. Additional comments on the procedure or method itself are provided in the **Notes** section.

## 2. MATERIALS

### 2.1 CABS-dock server methodology

The CABS-dock web server (freely available at http://biocomp.chem.uw.edu.pl/CABSdock/) provides an interface for the CABS-dock method for protein-peptide docking together with up-to-date documentation and benchmark examples [15]. Several illustrative examples of CABS-dock applications have also been described in [16, 21]. Here only the basic CABS-dock features are outlined. The CABS-dock server protocol is based on the CABS model [17-19, 22] for coarse-grained simulations of protein dynamics and protein structure prediction. The model employs a reduced representation of the protein chain (see Figure 1). The protein is represented with a set of pseudo-atoms: each residue is described by beads corresponding to the alpha carbon (**CA**), beta carbon (**B**) and side chain (**S**) (see Figure 1). To define the hydrogen bonds properly, an additional pseudo-atom representing the geometric center of the virtual CA-CA bond is also included. The knowledge-based force-field used for calculations was derived from statistical potentials based on known protein structures. Sampling of the conformation space is executed with a Replica Exchange Monte Carlo protocol. The CABS-dock docking procedure may be divided into four stages: (1) flexible docking based on the CABS model resulting in 10,000 models, (2) initial filtering resulting in 1000 models, (3) selection of 10 representative (top-ranked) models using structural clustering, (4) all-atom model reconstruction of 10 top-ranked models combined with local optimization of their structure. All those sets of models can be downloaded from the server web site for their visualization or analysis.

### 2.2 Running the CABS-dock server

A fully functional, up-to-date version of the CABS-dock method is available as an automated server accessible via standard internet browsers [15, 16]. No registration is required to use CABS-dock. To run the automated docking procedure on the CABS-dock server it is sufficient to provide:

- a 3D model of the protein receptor in the PDB format (the protein model should be provided in the standard PDB format; if the protein receptor structure is stored in the PDB databank, it is sufficient to provide its code only); for additional protein input hints see **Note 1**
- a peptide sequence and, optionally, peptide secondary structure in the one letter code; for additional peptide input hints see **Note 2**

The screenshots of the CABS-dock web server interface are shown in Figure 2. Docking results may be further improved by providing additional information about the protein complex (assigning regions of increased flexibility or excluded from docking, see **Note 3**).

## 3. METHODS

### 3.1 A case study of docking a peptide containing the LXXLL motif to PPARγ

This case study presents an example of CABS-dock docking performed with default server settings. The docking uses the protein receptor structure: peroxisome proliferator-activated receptor gamma (PPARγ) (PDB code of the unbound receptor form: 2HWQ) and the sequence of the peptide that contains the LXXLL motif of a cofactor protein crucial for the biological action of PPARγs [23]. Such a complex has been hypothesized to be responsible for the decoupling of insulin sensitization from adipogenesis in type-2 diabetes patients. The hypothesis was positively validated in vitro. A candidate for a partial PPARγ agonist was synthesized and crystallized by Burgermeister et. al [24] (PDB code of the complex: 2FVJ). The complex structure has been explored experimentally because of its potential for developing new therapies with fewer adverse effects on diabetes patients.

#### 3.1.1 Input and job submission

The "submit new job" form was completed in the following manner to attempt docking the peptide to the protein receptor:

1. Protein tab: "2HWQ:A" (this instructs the server to access the "A" chain of the 2HWQ structure). For additional hints regarding the input of a protein receptor structure see **Note 1**

2. Peptide tab: HKLVQLLTTT (this is the one-letter code sequence of the peptide containing the LXXLL docking motif of the protein cofactor). For additional hints regarding the input of a peptide sequence see **Note 2**

3. Optional tab:
   - Project name: "2HWQ:A tutorial" (used to identify the project in the server queue).
   - Peptide secondary structure: "CHHHHHHHCC"; this is the experimentally derived preferred secondary structure of the peptide. For additional hints see Note 2.
   - Additionally, an e-mail address may be provided. It will be used to notify the user on project status.

The run is started with the "Submit" button. The server will redirect the user to an auto-refreshing site with details on project status. Alternatively, it is possible to run the docking from the terminal command line using the following command (for further details on command-line job submission, see Section 3.4):

```
curl -H "Content-Type: application/json" -X POST -d '{"receptor_pdb_code":"2HWQ:A",
"ligand_seq":"HKLVQLLTTT","ligand_ss":"CHHHHHHCC","project_name":"2HWQ:A
tutorial", "email":"mail@host.com"}'
http://biocomp.chem.uw.edu.pl/CABSdock/REST/add_job/
```

### 3.1.2 Analysis of results

The results of docking may be either interactively viewed on the CABS-dock server or downloaded from the project site as a zipped folder with all the resulting files, see **Note 4**. The basic output provided by the CABS-dock server interface consists of 10 top ranked models (CABS-dock ranking is largely based on the outcome of structural clustering, for details see [15, 16]). The 10 top-ranked models are also stored in the zipped folder (in the form of PDB files named "model_(number).pdb"). The structures of models resulting from the docking performed for this case study are presented in Figure 3 together with the crystallographic structure of the protein (extracted from the 2FVJ PDB entry).

To analyze the quality of the resulting structures, calculation of RMSD values can be performed using for example VMD software [25]. A detailed tutorial for the VMD analysis of CABS-dock results is provided in the supplementary data in [16]. Our analysis below was performed using this tutorial to calculate RMSD values: first for the 10 top-ranked models, and second for the 10,000 models obtained in the CABS-dock simulation.

The RMSD values for the 10 top-ranked models to the crystal structure of the peptide (from the 2FVJ complex) are presented in Table 1. The lowest RMSD value of 1.29 Å was obtained for the model ranked as the sixth out of ten models (see Figure 3). Obviously, in the best case scenario the model with the lowest RMSD is ranked first. However, this is rarely the case as ranking the models is a very complex and yet unsolved problem (the scoring problem has been discussed in ref. [16]).

As briefly described in the Methods section, CABS-dock flexible docking produces a total of 10,000 models. For all these models RMSD values can also be easily calculated and plotted, for example against their CABS-energy values. Such analysis (showing for example whether the top-ranked models are also the lowest-RMSD models) is presented in Figure 4. The lowest RMSD model from all the 10,000 models has the accuracy of 1.00 Å and belongs to the set of near-native low energy models. As shown in Figure 4, apart from the low-energy and low-RMSD set of structures, there is also another low-energy set with RMSD around 9 Å. These structures also have their representatives in the set of 10 top-ranked models (i.e. models number 1, 4 and 5, see Table 1). The analysis of those cases proves that they fit into the appropriate binding site of the receptor. However, the peptide conformation differs from that of the crystallographic structure. With models 1 and 4 the C and N termini of the peptide are flipped, and in model 5 the peptide is bent and does not form a helix.

Please note that in this test case: (1) in several top-ranked models the actual binding site of the receptor protein was not found, and (2) the CABS-dock ranking procedure works relatively well (the lowest RMSD out of 10 top scored models is only slightly higher than out of 10,000 models). Obviously, these two points may be not satisfied in other docking cases and the detailed statistics of CABS-dock performance on a large benchmark data set is presented in detail in [15, 16]. Finally, it is important to note that the CABS-dock procedure is a Monte Carlo-based algorithm, which may lead to different results in different runs.

### 3.2 Simulation contact maps

#### 3.2.1 Maps: an overview

The CABS-dock server provides an additional tool for the analysis of docking simulations in the form of contact maps. These maps depict the frequencies at which a pair of receptor/peptide residues interacts during simulation. Such information may be utilized to investigate the binding mechanism and three-dimensional structures of intermediates that occur on complex formation (as presented in our study of the folding and binding of a disordered peptide [26]). It can also provide clues about potential mutation sites to alter the binding affinity of the peptide.

An archive with CABS-dock simulation contact maps (maps.tar) can be downloaded as part of the ZIP file with the results (see **Note 4**). The contact maps are both given in the MAP file format (txt files, see **Note 5**) and PNG images. The file names correspond to maps presenting contact frequencies of the following sets of models:

1. cluster_(number) – models classified to a particular cluster in structural clustering. Cluster numbering corresponds to model numbering (i.e. model_6.pdb is a representative model of the models grouped into the 6th cluster. The clusters are ranked according to their CABS-score)

2. trajectory_(number) – trajectory models (each of the trajectories contains 1000 models). Each CABS-dock job contains 10 trajectories

3. top1000 – top 1000 models selected after initial filtering

4. trajectory_all – all models from the 10 trajectories (10,000 models in total)

In the PNG images, contact frequencies are denoted by colors (example maps are presented in Section 3.2.2 below). Residue numbers and chain identifiers are marked on map borders. All the maps were derived from the distances of gravity centers of the side chains (in the CABS CG representation) and the contact cutoff was set to 4.5 Å.

#### 3.2.2 Example maps for docking a peptide containing the LXXLL motif to PPARγ

Because of their importance in further studies of the complex as well as potential significance in drug design, contact maps are one of the most informative results of the CABS-dock docking procedure. Most importantly, they may be used to predict residue-residue contacts that are crucial for the interaction, which for example can be subsequently used in peptide design.

According to experimental studies of the PPARγ-SRC-1 (a coactivator protein with the LXXLL motif) complex [27], the interaction site on the receptor protein is formed by the following residues: L468, L318, T297, Q314, L311, V315, K301 and E471. The docking LXXLL motif, which was experimentally determined to interact with PPARγs [23], is represented by residues 3-7 of the peptide used in the docking.

The contact maps for all the models (10,000 models in total, from the 10 trajectories) and cluster number 6 (the representative of this cluster is the lowest RMSD model from the top scored models) are presented in Figure 5. The maps show that the peptide residues comprising the motif, and the receptor residues creating native contacts in the crystallographic structure form the most persistent contacts during the CABS-dock docking simulation. The map prepared for all the simulation models (Figure 5a) shows that most of the (final) contacts are in the expected contact area.

Another informative way to visualize the engagement of particular residues in protein-peptide interaction during docking simulation is to prepare a histogram of residue contacts. The histogram can be prepared by summing up contact frequencies from the maps (available in MAP txt files) over the peptide residues. Two histograms for PPARγ receptor residues for all the models and cluster number 6 models are presented in Figure 6. The peaks found on both histograms correspond to residues crucial for the modeled interaction which form the interaction site of the receptor. The histogram for all the structures (Figure 6a) contains "background" noise resulting from peptide sampling of the receptor surface in search for the best binding position. Some of those interactions are more persistent (e.g., residue 259) and may take part in intermediate complex formation while most of them are likely to be accidental. Although not all of the expected contacts were present in the resulting structures, it is clear that the most important interactions are well preserved and visible in the models. It is also possible that further all-atom refinement of the complex structure may lead to enhancement of the interaction site details that are not clear in the CG representation (see Section 3.3 below).

Finally, note that the contact map analysis of the folding and binding of a disordered peptide (simulated using CABS-dock methodology) has also been presented in [26].

### 3.3 CABS-dock: possible applications and future advances

It is expected that computational techniques will play an important role in the rational design of peptide therapeutics [3]. Peptides make very promising candidates for drugs as they can adopt multiple shapes and various chemical features through careful design. Moreover, the design and synthesis of peptide drugs is relatively simple, so large libraries of peptides may be easily scanned to look for optimal peptide design.

The CABS-dock server may be used as an initial docking tool in a multistage docking procedure. Perhaps the most straightforward CABS-dock application is to use it as a tool for determining the initial structure(s) of a protein-peptide complex that may be used as an input for further refinement by local docking methods [9-14]. As shown before on a large protein-peptide benchmark dataset [15, 16], for the majority of cases CABS-dock produced models with high or medium accuracy (for example sufficient for structure refinement by Rosetta FlexPepDock [10, 12]). Another conclusion from the benchmark analysis was that CABS-dock accuracy can be significantly improved by its combination with exact scoring methods. By default, top-ranked models produced in the CABS-dock procedure are reconstructed to all-atom representation and refined using MODELLER [28] procedures and ranked by the DOPE score [29]. Since the reconstruction and the final all-atom refinement may significantly alter the quality of models, other techniques (better suited for the reconstruction and optimization of CABS-dock coarse-grained models) may be highly useful.

Future CABS-dock improvements also include its integration with methods for prediction of peptide binding-sites [5-8] or extending the CABS-dock functionality to user guided docking (by providing a possibility of pointing residues that belong to the binding site). Narrowing the conformation space to the selected neighborhood should result in the better sampling of near-native states, and thus in increasing the chances for building high accuracy models. Virtually any structural information may be utilized by CABS-dock as distance restraints or filters. Therefore, CABS-dock is well suited to be integrated as an efficient sampling tool with computational pipelines for modeling protein-peptide interactions, including methods for *de novo* peptide design [30, 31] or template-based docking [32].

Finally, CABS-dock could be used in hierarchical protein-protein docking protocols composed of three modeling steps:
(1) Reduction of the protein-protein docking problem to protein-peptide docking. This starts from the arbitrary selection of the receptor protein and bound protein, followed by the identification of 'hot segment(s)' of the bound protein, i.e. a short epitope that contributes the most to the protein-protein interaction [33, 34]
(2) CABS-dock docking of 'hot segment(s)' [33, 34], i.e. peptide(s)
(3) Reconstruction and adjustment of the remaining receptor structure to the docked peptide-like fragment

Peptide-like 'hot segment(s)' can be of various length and can represent more than one fragment of the original structure, provided that they can be realistically replaced by a continuous peptide chain. In the context of the potential application of CABS-dock in protein-protein docking described above, one can also easily design a simple sequential procedure for the efficient modeling of amyloid aggregation.

### 3.4 Running CABS-dock from the command line

Except for using the web interface (available at http://biocomp.chem.uw.edu.pl/CABSdock/), the CABS-dock server can also be operated from the command line or scripts using REST-full service. This option is recommended for handling multiple jobs by users experienced in Bash and python scripting.

#### 3.4.1  Submitting a job with the PDB code of a protein receptor

To submit a job for a chosen protein, e.g. 2GB1, and a peptide sequence, e.g. SFDG, with default parameters, the following command or python script should be run:

- **command line:**
```
curl -H "Content-Type: application/json" -X POST -d
  '{"receptor_pdb_code":"2GB1",
  "ligand_seq":"SFGD"}'  http://biocomp.chem.uw.edu.pl/CABSdock/REST/add_job/
```

- **python script:**
```
import requests
import json
url = ' http://biocomp.chem.uw.edu.pl/CABSdock/REST/add_job/'
data = {
   "receptor_pdb_code": "2GB1",
   "ligand_seq": "SFGD",
}
response = requests.post(url, data=data)
```

The PDB file corresponding to "receptor_pdb_code" will be automatically downloaded from the PDB database. On success, a job identifier assigned to the submitted job "jid" will be returned. Jid will be used as a query for the job status and results later on. Otherwise, for example if the pdb code doesn't exist or input data don't fulfill requirements, error will be signaled.

### 3.4.2 Submitting a job with a user-provided PDB file

Instead of the PDB code, a PDB file can be attached to the query in the following ways:

- **command line:**

```
curl -X POST -F data='{"ligand_seq":"SFGD"}' -F
   file=@path_to_pdb_file.pdb  http://biocomp.chem.uw.edu.pl/CABSdock/REST/add_j
   ob/
```

- **python script:**

```
import requests
import json
url = ' http://biocomp.chem.uw.edu.pl/CABSdock/REST/add_job/'
files = {'file': open('path_to_pdb_file.pdb')}
data = {
    "ligand_seq": "SFGD",
}

response = requests.post(url, files=files, data=data)
```

### 3.4.3 Overriding default parameters

To override default parameters, additional options may be posted, i.e.:

- **command line:**
```
curl -H "Content-Type: application/json" -X POST -d
   '{"receptor_pdb_code":"2IV9",
   "ligand_seq":"SFGD","project_name":"my_project1", "email":"mail@host.com",
   "ligand_ss":"CCHHC", "simulation_cycles":"100", "show_job":True,
   "excluded_regions":[{"start":"100","end":"340","chain":"A"}],
   "flexible_regions":[{"start":"101","end":"202","chain":"B","flexibility":"ful
   l"}]}'  http://biocomp.chem.uw.edu.pl/CABSdock/REST/add_job/
```

- **python script:**
```
import requests
import json
url = ' http://biocomp.chem.uw.edu.pl/CABSdock/REST/add_job/'
files = {'file': open('your_PDB_file.pdb')} #or use PDB code in var data
data = {
    "receptor_pdb_code": "2IV9", #or use PDB file in var files
    "ligand_seq": "SFGD",
    "email": "mail@host.com",
    "show": True,
    "project_name":"my_project1",
    "excluded_regions":[
        {
            "start": "1000",
            "end": "2000",
            "chain": "A"
        }
    ],
    "flexible_regions":[
        {
            "start": "101",
```

```
            "end": "202",
            "chain": "A",
            "flexibility": "full"
        },
        {
            "start": "300",
            "end": "370",
            "chain": "B",
            "flexibility": "moderate"
        },
    ]
}
response = requests.post(url, files=files, data=data) #request with file
#response = requests.post(url, data=data) # request without file
```

**Additional parameters include:**
- project_name – name of the project used for job identification, i.e. in the queue
- email – email used to inform the user about job progress
- ligand_ss – ligand secondary structure
- simulation_cycles – number of simulation cycles: the default is 100 and the maximum is 200
- show_job – boolean value (True or False) indicating whether to show a job on the queue page
- excluded_regions – array of excluded regions. Each excluded region represents a selected receptor residue that is unlikely to interact with the peptide and should contain the following fields: start position, end position and chain
- flexible_regions – array of flexible regions. The flexibility of the region is changed by removing distance restraints that keep the receptor structure in a near native conformation. Each element of the array contains start position, end position, chain and flexibility. Flexibility can be either full or moderate

### 3.4.4 Getting job status

To check the status of a job, a job identifier ("jid") should be provided:

- **command line:**
```
curl -I "http://biocomp.chem.uw.edu.pl/CABSdock/REST/status/somejobidentifier"
```

- **python script:**

```
import requests
import json
url = 'http://biocomp.chem.uw.edu.pl/CABSdock/REST/status/somejobidentifier'
response = requests.post(url)
```

As a result, one of the following statuses will be returned:
- `done` – job is finished and the results are ready
- `pending / running / pre_quere` – job is in progress
- `error` – the job identifier doesn't exist

More detailed information about the job can be obtained by running:

- **command line:**
```
curl -I "http://biocomp.chem.uw.edu.pl/CABSdock/REST/job_info/somejobidentifier"
```

- **python script:**
```
import requests
import json
url = 'http://biocomp.chem.uw.edu.pl/CABSdock/REST/job_info/somejobidentifier'
response = requests.post(url)
```

Additional information includes job configuration which was provided on submission and more details about the status. The following fields will be listed in the result:

- del – job results will be kept on the server until this date
- excluded – list of excluded regions sent on job submission
- flexible – list of flexible regions sent on job submission
- ligand_sequence – ligand sequence sent on job submission
- ligand_ss – ligand secondary structure sent on job submission
- project_name – name assigned to the project on job submission
- receptor_sequence – receptor sequence sent on job submission
- ss_psipred – secondary structure predicted by psipred
- status – one of the possible job statuses as described in the section Getting job status
- status_change – time of last status change

### 3.4.5 Getting job results: essential information

Essential information for each model includes:
- Average RMSD
- Max RMSD
- Cluster density
- Number of elements
- Model data
- Information about submitted data

See the next chapter for more information.

To obtain essential information, the job identifier ("jid") must be provided:

- **command line:**
```
curl -i "http://biocomp.chem.uw.edu.pl/CABSdock/REST/get_job/somejobidentifier"
```

We strongly recommend that curl with compression should be sent:
```
curl -i -H 'Accept-Encoding: gzip,deflate'
    "http://biocomp.chem.uw.edu.pl/CABSdock/REST/get_job/somejobidentifier"
```

- **python script:**
```
import requests
import json
url = 'http://biocomp.chem.uw.edu.pl/CABSdock/REST/get_job/somejobidentifier'
response = requests.post(url)
```

Optional parameters for filtering the results can be attached to the query. The parameters must specify the attribute used for filtering ("value") and the allowed range of values for the attribute ("min" and "max"). The following attributes can be used for filtering:
- density – cluster density
- rmsd – average RMSD
- maxrmsd – maximum RMSD
- counts – number of elements in a cluster

Exemplary use of filtering:

- **command line:**
```
curl -i -X POST -d
   '{"filter":"density","min":"10","max":"20"}'  "http://biocomp.chem.uw.edu.pl/
   CABSdock/REST/get_job/somejobidentifier"
```

- **python script:**
```
import requests
import json
url = 'http://biocomp.chem.uw.edu.pl/CABSdock/REST/get_job/somejobidentifier'
data = {
       "value":"rmsd",
       "min":"5",
       "max":"12"
       }
response = requests.post(url, data=data)
```

### 3.4.5 Getting job results: all information

All information for each model includes:
- Average RMSD
- Max RMSD
- Cluster density
- Number of elements
- Model data
- Information about submitted data

and additionally:
- Cluster data

To get cluster data or trajectory data, see the next sections.

To obtain all information, the job identifier ("jid") must be provided:
- **command line:**
```
curl -i
   "http://biocomp.chem.uw.edu.pl/CABSdock/REST/get_job_all/somejobidentifier"
```

We strongly recommend that curl with compression should be sent:
```
curl -i -H 'Accept-Encoding: gzip,deflate'
   "http://biocomp.chem.uw.edu.pl/CABSdock/REST/get_job_all/somejobidentifier"
```

- **python script:**
```
import requests
import json
```

```
url = 
'http://biocomp.chem.uw.edu.pl/CABSdock/REST/get_job_all/somejobidentifier'
response = requests.post(url)
```

Additional filtering can be applied to the query as described in the previous section

### 3.4.6 Getting cluster information

To get information about a chosen cluster, the job identifier together with the cluster number corresponding to the model number should be submitted:

- **command line:**
```
curl -i
   "http://biocomp.chem.uw.edu.pl/CABSdock/REST/get_cluster/somejobidentifier/clusterNumber"
```

We strongly recommend that curl with compression should be sent:
```
curl -i -H 'Accept-Encoding: gzip,deflate'
   "http://biocomp.chem.uw.edu.pl/CABSdock/REST/get_cluster/somejobidentifier/clusterNumber"
```

- **python script:**
```
import requests
import json
url = 
'http://biocomp.chem.uw.edu.pl/CABSdock/REST/get_cluster/somejobidentifier/clusterNumber'
response = requests.post(url)
```

The cluster number must be in the range [1, 10]. As a result, cluster data and additional information about the cluster (average and maximum RMSD, cluster density and number of elements) will be returned.

### 3.4.7 Getting trajectory information

Trajectory data can be obtained by sending a query with the attached job identifier and model number in the range [1,10]:
- **command line:**
```
curl -i
   "http://biocomp.chem.uw.edu.pl/CABSdock/REST/get_trajectory/somejobidentifier/modelNumber"
```

We strongly recommend that curl with compression should be sent:
```
curl -i -H 'Accept-Encoding: gzip,deflate'
   "http://biocomp.chem.uw.edu.pl/CABSdock/REST/get_trajectory/somejobidentifier/modelNumber"
```

- **python script:**
```
import requests
   import json
```

```
url =
'http://biocomp.chem.uw.edu.pl/CABSdock/REST/get_trajectory/somejobidentifier
/modelNumber'
response = requests.post(url)
```

Additionally, a section of the trajectory model can be selected by:

- **command line:**
```
curl -i
   "http://biocomp.chem.uw.edu.pl/CABSdock/REST/get_selected_trajectory/somejobi
   dentifier/modelNumber/start/end"
```

- **python script:**
```
import requests
   import json
   url =
'http://biocomp.chem.uw.edu.pl/CABSdock/REST/get_selected_trajectory/somejobi
   dentifier/modelNumber/start/end'
   response = requests.post(url)
```

### 3.4.8 Examples

*Example 1 (default settings)*

The first example shows how to submit a job with the following data:
- Peptide sequence: SSRFESLFAG
- Peptide secondary structure: CHHHHHHHHC
- Receptor input structure: PDB ID, 2AM9, crystal structure of the human androgen receptor in the unbound form

and the default CABS-dock server settings.

- **command line:**

```
curl -H "Content-Type: application/json" -X POST -d
   '{"receptor_pdb_code":"2AM9", "ligand_seq":"SSRFESLFAG",
   "ligand_ss":"CHHHHHHHHC"}'
   "http://biocomp.chem.uw.edu.pl/CABSdock/REST/add_job/"
```

- **python script:**
```
import requests
import json
url = 'http://biocomp.chem.uw.edu.pl/CABSdock/REST/add_job/'
data = {
   "receptor_pdb_code": "2AM9"
   "ligand_seq": "SSRFESLFAG",
   "ligand_ss": "CHHHHHHHHC"
}
response = requests.post(url, data=data)
```

*Example 2 (increasing the flexibility of selected receptor fragments)*

The second example shows how to increase the flexibility of selected receptor fragments.

For each selected residue, one of two settings of flexibility (moderate or full) can be set. Technically, this is achieved by changing the default distance restraints used to keep the receptor structure near to the input conformation. The assignment of moderate flexibility decreases the strength of restrains, while full flexibility assignment removes all the restraints imposed on the selected residue.

Data used in the example:
- Peptide sequence: HPQFEK
- Peptide secondary structure: CHHHCC
- Receptor input structure: PDB ID: 2RTM, crystal structure of biotin binding protein in the unbound form

Additional options:
- Using the CABS-dock "Mark flexible regions" option, 10 residues (45 to 54) forming the flexible loop are selected and the fully flexible☐ setting is assigned to those residues.

Important : Numbering in the PDB format must be used.

- **command line:**
```
curl -H "Content-Type: application/json" -X POST -d 
   '{"receptor_pdb_code":"2RTM", "ligand_seq":"HPQFEK", "ligand_ss":"CHHHCC",
   "flexible_regions":[{"start":"45","end":"54","chain":"A","flexibility":"full"
   }]}' http://biocomp.chem.uw.edu.pl/CABSdock/REST/add_job/
```

- **python script:**
```
import requests
import json

url = 'http://biocomp.chem.uw.edu.pl/CABSdock/REST/add_job/'
data = {
   "receptor_pdb_code": "2RTM"
   "ligand_seq": "HPQFEK",
   "ligand_ss": "CHHHCC"
   "flexible_regions":[
       {
           "start": "45",
           "end": "54",
           "chain": "A",
           "flexibility": "full"
       }
   ]
}
response = requests.post(url, data=data)
print response.text
```

*Example 3 (excluding binding modes from docking search)*

The third example focuses on excluding binding modes form docking search. In the default mode, CABS-dock allows peptides to explore the entire receptor surface. However, in certain modeling cases it is known that some parts of the protein are not accessible (for example due to binding to other proteins) and therefore it could be useful to exclude these regions from the search.

Data used in the example:
- Peptide sequence: PQQATDD

- Peptide secondary structure: CEECCCC
- Receptor input structure: PDB ID: 1CZY:C, tumor necrosis factor receptor associated protein 2 in the unbound form

Additional options:
- 1CZY protein is a trimer and 1CZY:C forms contacts with 1CZY:A (according to the http://ligin.weizmann.ac.il/cma/ server for the analysis of protein-protein interfaces). Therefore the residues in the C chain (the input protein) listed above which are responsible for contacts with A and B chains can be excluded from the docking search using the CABS-dock "Exclude regions" option.

- **command line:**

```
curl -H "Content-Type: application/json" -X POST -d
    '{"receptor_pdb_code":"1CZY:C", "ligand_seq":"PQQATDD",
    "ligand_ss":"CEECCCC", "excluded_regions":[
    {"start":"334","end":"335","chain":"C"},
    {"start":"338","end":"338","chain":"C"},
    {"start":"341","end":"342","chain":"C"},
    {"start":"345","end":"345","chain":"C"},
    {"start":"350","end":"350","chain":"C"},
    {"start":"385","end":"386","chain":"C"},
    {"start":"416","end":"418","chain":"C"},
    {"start":"420","end":"421","chain":"C"},
    {"start":"458","end":"458","chain":"C"} ]}'
    http://biocomp.chem.uw.edu.pl/CABSdock/REST/add_job/
```

- **python script:**

```
import requests
import json

url = 'http://biocomp.chem.uw.edu.pl/CABSdock/REST/add_job/'
files = {'file': open('your_PDB_file.pdb')} #or use PDB code in var data
data = {
   "receptor_pdb_code": "1CZY:C"
   "ligand_seq": "PQQATDD",
   "ligand_ss": "CEECCCC"
   "excluded_regions":[
       {
           "start": "334",
           "end": "335",
           "chain": "C",
       },
       {
           "start": "338",
           "end": "338",
           "chain": "C",
       },
       {
           "start": "341",
           "end": "342",
           "chain": "C",
       },
       {
           "start": "345",
           "end": "345",
```

```
                    "chain": "C",
                },
                {
                    "start": "350",
                    "end": "350",
                    "chain": "C",
                },
                {
                    "start": "385",
                    "end": "386",
                    "chain": "C",
                },
                {
                    "start": "416",
                    "end": "418",
                    "chain": "C",
                },
                {
                    "start": "420",
                    "end": "421",
                    "chain": "C",
                },
                {
                    "start": "458",
                    "end": "458",
                    "chain": "C",
                }
        ]
        }
response = requests.post(url, data=data)
```

## 4 NOTES

1. The CABS-dock server requires a user provided protein receptor structure in the PDB format or the PDB code of the receptor (the file will be automatically downloaded to the server from the PDB database). The chain of the protein receptor must be shorter than 500 amino acids. The backbone must be complete; however side chain atoms may be missing. Any non-standard amino acids in the protein receptor will be changed to their standard counterparts.

2. The peptide sequence input must be 4-30 amino acids in length and consist of standard amino acids only. It is also possible to provide the secondary structure of the peptide in the standard one-letter code (C – coil, H – helix, E – extended) using the "Optional" tab (if not, the secondary structure will be predicted with PsiPred). The structure may be experimentally derived or based on any sequence-based prediction method. Please note that "overprediction" of regular structures (H, E) was shown to be more likely to give incorrect results of docking than their underprediction. If the secondary structure is not known, it is better to supply it as a list of "C" (coil assignments). More information on how the secondary structure information is used in the simulations is provided in reference [35].

3. On top of standard input settings the CABS-dock server provides an advanced input

panel which enables additional features to tailor simulation conditions to the user's needs. These features include: (1) Custom adjusted run time: the user is allowed to lengthen the simulation run time, which may save time in case of small complexes or lead to better results for large complexes, where the standard setting may be insufficient to cover the whole conformational space. (2) Selection of flexible regions of the receptor: the user may mark some of the residues of the receptor to be granted more conformational flexibility than in the standard settings. By default receptor residues are flexible, but limited to only near-native conformations, which is suitable for most docking applications. Additional flexibility may be adjusted to semi- or full flexible to model more accurately regions believed to change their conformation on peptide binding. (3) Exclusion from sampling the receptor regions unlikely to be involved in peptide binding: the user may select some of the receptor residues believed not to take part in peptide binding. This feature is useful when the receptor molecule contains more than one binding spot and only one needs to be investigated (i.e. in receptors containing dimerization sites) or when part of the receptor is inaccessible to the peptide in vivo (i.e. receptors embedded in the membrane). Illustrative examples of using these advanced features are provided in [16].

4. All CABS-dock results can be downloaded in a single ZIP archive file available from the "Docking predictions results" tab. The ZIP archive file contains the simulation trajectories, clusters of models and the top ranked models (representatives of the clusters). All the provided structures are in PDB format files and the top ranked models are provided in all-atom resolution. The trajectories and cluster model coordinates are provided in C-alpha representation only. The ZIP archive also contains simulation contact maps (discussed in Section 3.2).

5. The contact maps are stored as PNG figures and MAP files. The MAP file is a text file (txt) that consists of three columns: the first two list the residues of the protein receptor and the peptide, respectively. In each row, the third column gives the frequency of the contact between the residues in first two columns. An example fragment of a MAP file format is presented below:

```
...
A224   C7    0.0117647
A224   C8    0.0117647
A224   C9    0
A225   C1    0
A225   C10   0.0117647
A225   C2    0
A225   C3    0
A225   C4    0
A225   C5    0
A225   C6    0
A225   C7    0.0117647
...
```

Each of the residues in the receptor protein is paired with each residue of the peptide, so the number of rows in the file is (number of protein residues)*(number of peptide residues).

**Acknowledgments**

The authors acknowledge support from the National Science Center grant [MAESTRO 2014/14/A/ST6/00088]
**References**

1. Tsomaia, N., *Peptide therapeutics: targeting the undruggable space.* Eur J Med Chem, 2015. **94**: p. 459-70.
2. Fosgerau, K. and T. Hoffmann, *Peptide therapeutics: current status and future directions.* Drug Discov Today, 2015. **20**(1): p. 122-8.
3. Diller, D.J., J. Swanson, A.S. Bayden, M. Jarosinski, and J. Audie, *Rational, computer-enabled peptide drug design: principles, methods, applications and future directions.* Future Med Chem, 2015. **7**(16): p. 2173-93.
4. London, N., B. Raveh, and O. Schueler-Furman, *Peptide docking and structure-based characterization of peptide binding: from knowledge to know-how.* Curr Opin Struct Biol, 2013. **23**(6): p. 894-902.
5. Yan, C. and X. Zou, *Predicting peptide binding sites on protein surfaces by clustering chemical interactions.* J Comput Chem, 2015. **36**(1): p. 49-61.
6. Verschueren, E., P. Vanhee, F. Rousseau, J. Schymkowitz, and L. Serrano, *Protein-peptide complex prediction through fragment interaction patterns.* Structure, 2013. **21**(5): p. 789-97.
7. Saladin, A., J. Rey, P. Thevenet, M. Zacharias, G. Moroy, and P. Tuffery, *PEP-SiteFinder: a tool for the blind identification of peptide binding sites on protein surfaces.* Nucleic Acids Res, 2014. **42**(Web Server issue): p. W221-6.
8. Lavi, A., C.H. Ngan, D. Movshovitz-Attias, T. Bohnuud, C. Yueh, D. Beglov, O. Schueler-Furman, and D. Kozakov, *Detection of peptide-binding sites on protein surfaces: the first step toward the modeling and targeting of peptide-mediated interactions.* Proteins, 2013. **81**(12): p. 2096-105.
9. Antes, I., *DynaDock: A new molecular dynamics-based algorithm for protein-peptide docking including receptor flexibility.* Proteins, 2010. **78**(5): p. 1084-104.
10. London, N., B. Raveh, E. Cohen, G. Fathi, and O. Schueler-Furman, *Rosetta FlexPepDock web server--high resolution modeling of peptide-protein interactions.* Nucleic Acids Res, 2011. **39**(Web Server issue): p. W249-53.
11. Trellet, M., A.S. Melquiond, and A.M. Bonvin, *A unified conformational selection and induced fit approach to protein-peptide docking.* PLoS One, 2013. **8**(3): p. e58769.
12. Raveh, B., N. London, L. Zimmerman, and O. Schueler-Furman, *Rosetta FlexPepDock ab-initio: simultaneous folding, docking and refinement of peptides onto their receptors.* PLoS One, 2011. **6**(4): p. e18934.
13. Trellet, M., A.S. Melquiond, and A.M. Bonvin, *Information-driven modeling of protein-peptide complexes.* Methods Mol Biol, 2015. **1268**: p. 221-39.
14. Donsky, E. and H.J. Wolfson, *PepCrawler: a fast RRT-based algorithm for high-resolution refinement and binding affinity estimation of peptide inhibitors.* Bioinformatics, 2011. **27**(20): p. 2836-42.
15. Kurcinski, M., M. Jamroz, M. Blaszczyk, A. Kolinski, and S. Kmiecik, *CABS-dock web server for the flexible docking of peptides to proteins without prior knowledge of the binding site.* Nucleic Acids Res, 2015. **43**(W1): p. W419-W424.
16. Blaszczyk, M., M. Kurcinski, M. Kouza, L. Wieteska, A. Debinski, A. Kolinski, and S. Kmiecik, *Modeling of protein-peptide interactions using the CABS-dock web server for binding site search*

# Figure captions

**Figure 1. Representation of protein and peptide chains in CABS-dock.** CABS-dock uses all-atom and coarse-grained modeling tools merged with procedures enabling transition between both resolutions. The figure shows comparison between all-atom (left) and CABS coarse-grained representation (right) for an example 4-residue peptide. In the CABS model, a single residue is represented by 2 atoms (alpha and beta carbon, colored in black) and 2 pseudo-atoms (side chain, colored in orange, and center of the peptide bond, colored in green).

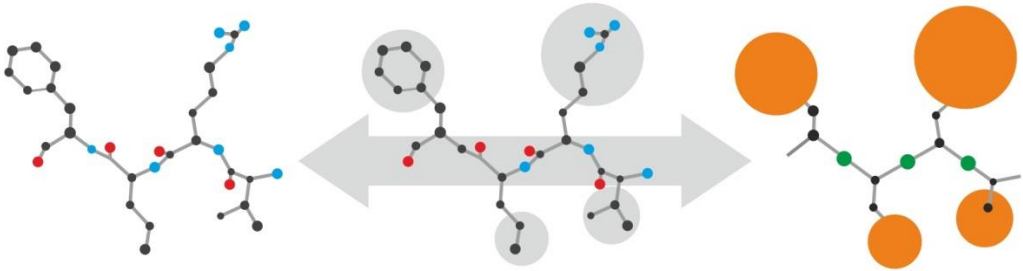

**Figure 2. Screenshots of the CABS-dock server.** The figure shows the main page input panel (a) and example output panels (b, c, d). The buttons to be selected to see these panels are marked by red rectangles and arrows.

**Figure 3. Steps of the CABS-dock docking procedure illustrated by peptide-PPARγ docking.** (a) Ten random peptide structures placed in random positions around the PPARγ structure. (b) 10,000 peptide structures generated in the CABS-dock docking simulation. (c) 1000 models filtered from the previous set. (d) 10 top-ranked models (according to the structural clustering analysis) resulting from the docking. The close-up frame shows the best fitting model (RMSD value of 1.29 Å) out of the 10 top-ranked models. The peptide models resulting from docking are shown in orange, the crystallographic peptide structure is shown in yellow and the protein receptor is represented by its surface with elements of the secondary structure visible.

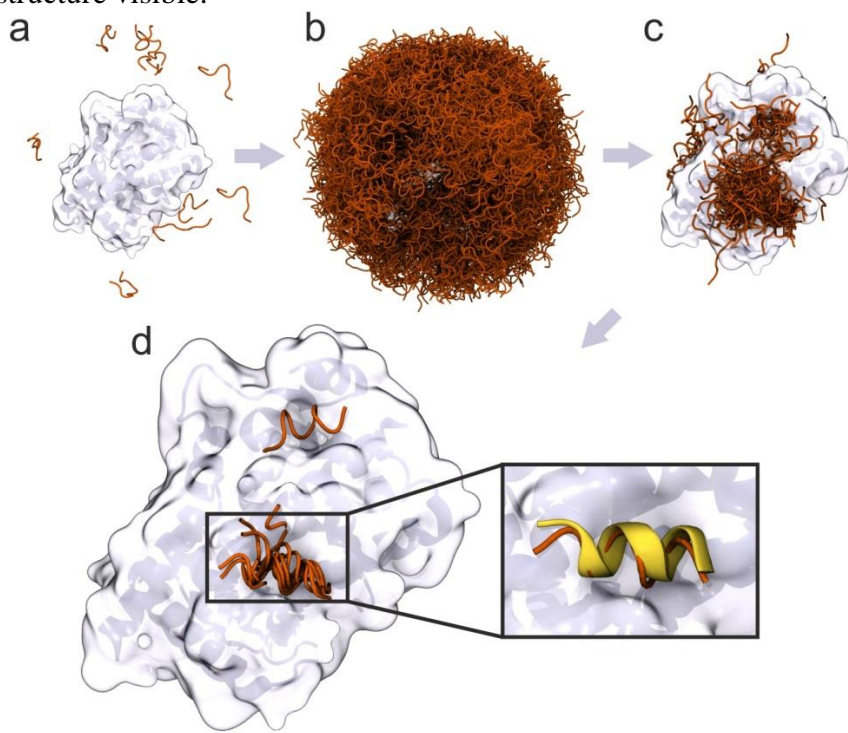

**Figure 4. CABS-energy and RMSD values for all (10,000) models obtained in peptide-PPARγ docking**. The colors of the dots represent 10 trajectories of a single docking simulation.

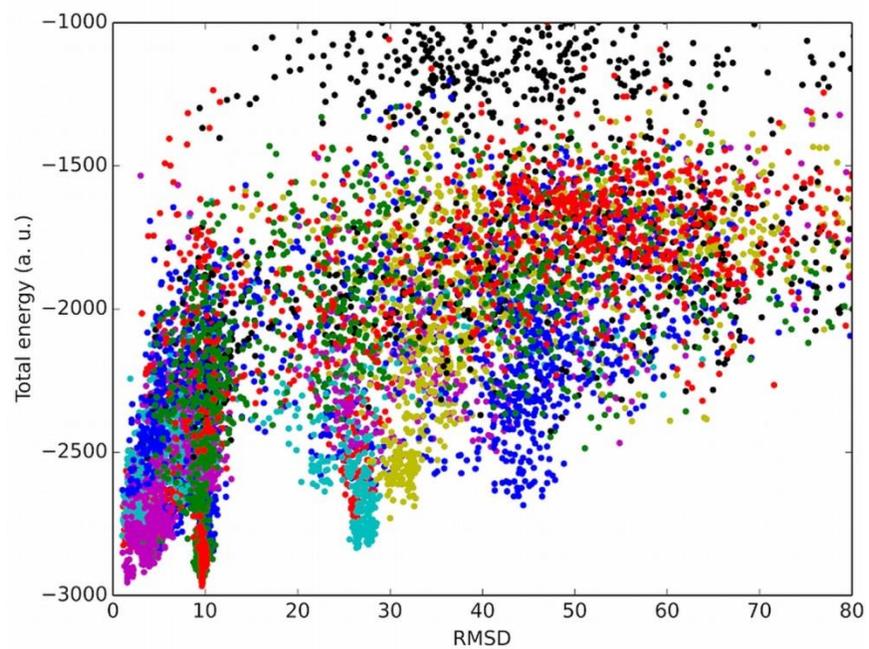

**Figure 5. Protein-peptide contact maps from peptide-PPARγ docking.** (a) Contact map for all 10,000 models. (b) Contact map for the models from cluster number 6 whose representative was the model best fitting the experimental structure. The columns represent amino acids of the receptor, and the rows are amino acids of the peptide. The residues reported in the literature to form the interaction site of the complex are marked in green. Contact frequencies are marked according to the color maps below each of the maps. The maps were divided into four elements for clarity of presentation.

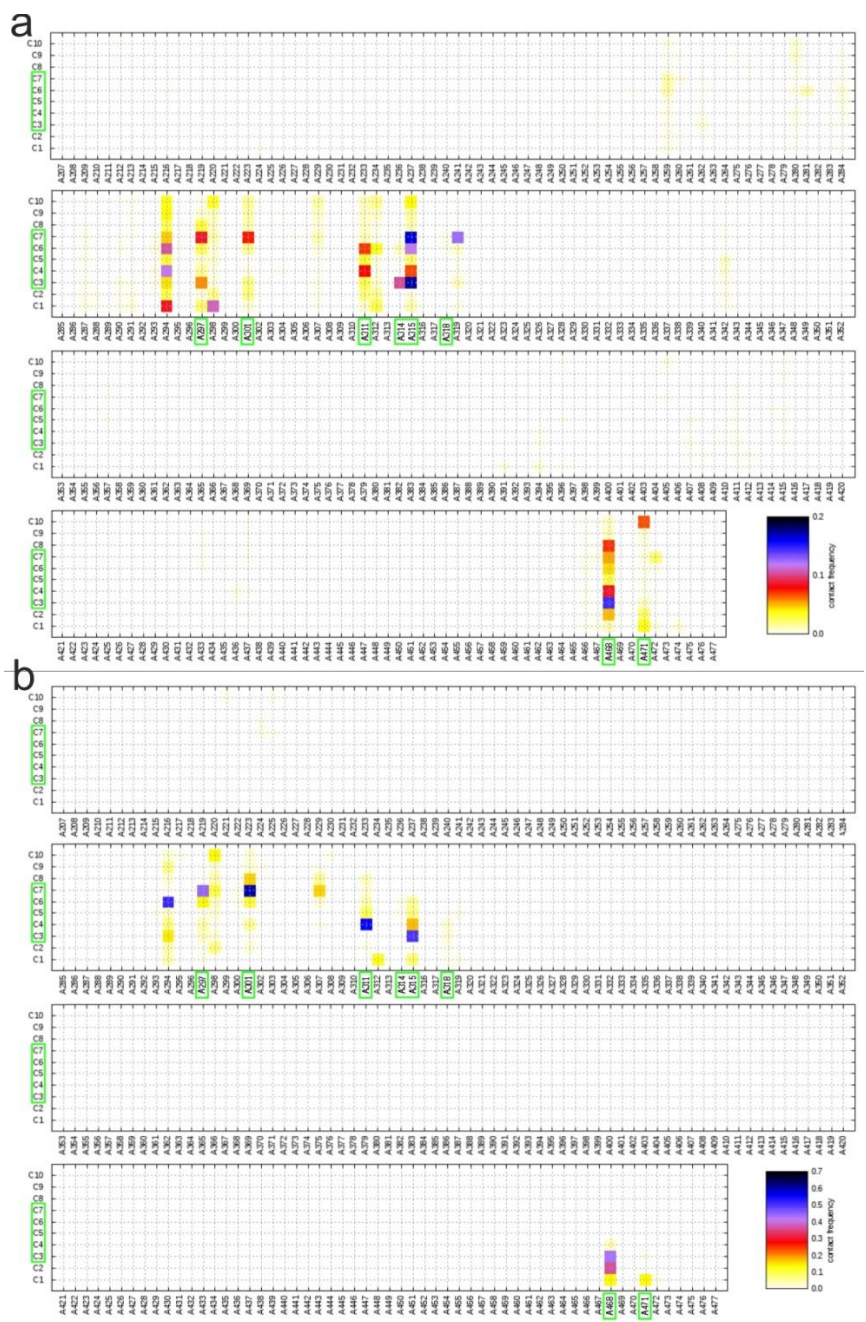

**Figure 6. Histograms of protein-peptide contacts from peptide-PPARγ docking**. The normalized histograms show frequencies of contact for each of the receptor residues with the peptide: (a) for all 10,000 models, (b) for the models from cluster 6 whose representative was the best fitting model. The green markers represent residues that were reported in the literature to form a pocket for the LXXLL-peptide motif on the surface of PPARγ.

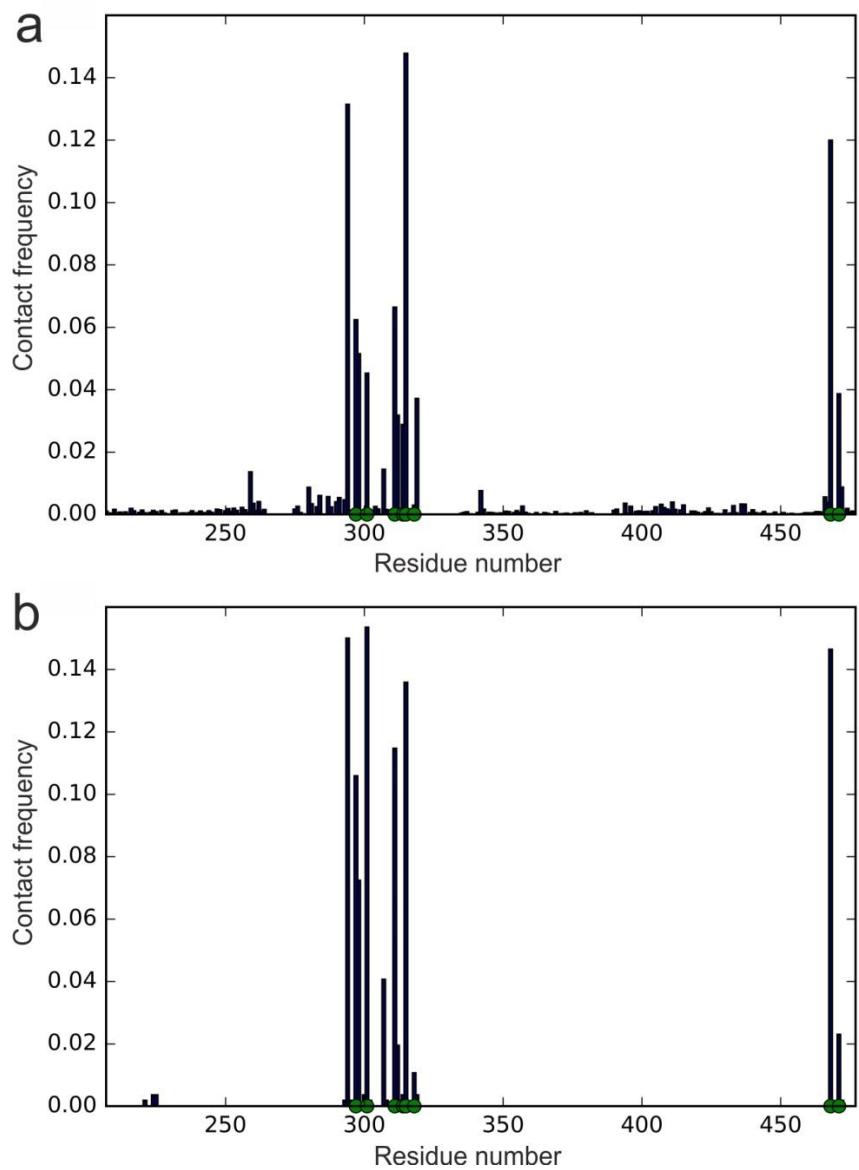

**Table captions**

**Table 1. The RMSD values of 10 top-ranked models to the crystallographic structure.** The entry for the best fitting structure is marked in bold. The provided RMSD values are root mean square deviations calculated on the peptides after superposition of the receptor molecules.

| Index of top ranked models | RMSD value |
|---|---|
| 1 | 9.705 |
| 2 | 3.444 |
| 3 | 3.801 |
| 4 | 9.987 |
| 5 | 9.242 |
| **6** | **1.290** |
| 7 | 3.610 |
| 8 | 1.778 |
| 9 | 31.353 |
| 10 | 26.036 |